\documentclass[showpacs]{revtex4}
\usepackage{amsmath}
\usepackage{amssymb}
\usepackage{graphicx}
\usepackage{color}
\usepackage{dcolumn}

\begin{document}
\newcolumntype{.}{D{.}{.}{1.6}}

\title{
Magnetic fields in anisotropic relativistic stars
}

\author{{\bf Vladimir~Folomeev$^{1}$}}
\email[{\it Email:}]{vfolomeev@mail.ru}
\author{{\bf Vladimir Dzhunushaliev$^{1,2,3,4}$}}
\email[{\it Email:}]{v.dzhunushaliev@gmail.com}
\affiliation{
$^1$Institute of Physicotechnical Problems and Material Science of the NAS
of the
Kyrgyz Republic, 265 a, Chui Street, Bishkek, 720071,  Kyrgyzstan\\
$^2$Department of Theoretical and Nuclear Physics,
Al-Farabi Kazakh National University, Almaty 050040, Kazakhstan\\
$^3$Institute of Experimental and Theoretical Physics,
Al-Farabi Kazakh National University, Almaty 050040, Kazakhstan
\\
$^4$Institute for Basic Research,
Eurasian National University,
Astana 010008, Kazakhstan
}

\begin{abstract}
Relativistic, spherically symmetric configurations consisting of a gravitating magnetized anisotropic fluid are studied.
For such configurations, we obtain static equilibrium solutions with an axisymmetric, poloidal magnetic field
produced by toroidal electric currents.
The presence of such a field results in small deviations of the shape of the configuration from spherical symmetry.
This in turn leads to the modification of an equation for the current and correspondingly
to changes in the structure of the internal magnetic field for the systems supported by the anisotropic fluid,
in contrast to the case of an isotropic fluid, where such deviations do not affect the magnetic field.
\end{abstract}


\maketitle

\section{Introduction}

Magnetic fields are widely believed to play an important role in modeling compact, strongly gravitating objects~-- neutron stars.
The measured surface magnetic fields of such objects are in the range of $10^{11}-10^{13}$~G (for the so-called ``classical pulsars'')
and can reach values of the order of $10^{15}$~G for magnetars.
Considerable research has been devoted to stellar magnetic fields
(see, e.g., the book~\cite{Mest}  on stellar magnetism in general and some works~\cite{Bocquet:1995je,magn_stars}
on particular questions
of the structure of strongly magnetized configurations, and also references therein).
The studies show that strong magnetic fields affect considerably the interior structure of neutron stars
and determine their evolution in time (the magnetic dipole radiation, the deformation due to the magnetic stress, etc.).

In modeling magnetized neutron stars, neutron matter is usually described as being an isotropic perfect fluid,
i.e., a fluid obeying Pascal's law. However, such a description cannot be considered as completely satisfactory,
since at extreme densities of matter and magnetic fields typical for the interior regions of neutron stars,
it is already necessary to take into account the interaction of matter with ultra-strong magnetic fields.
The presence of such fields
may lead to the appearance of unequal
principal stresses in the neutron fluid (see Ref.~\cite{anis_magn} and references therein);
i.e., the fluid will become anisotropic.
(See, however, Ref.~\cite{anis_magn_diss} for further discussion of the question of whether the hydrodynamic pressure  is isotropic or anisotropic
in the presence of magnetic fields.)
Among the other possible reasons for the appearance of the
anisotropy in superdense matter might be
the nuclear interactions~\cite{Rud1972},  pion condensation~\cite{Saw1972},
 some kinds of phase transitions~\cite{Sok1980}, or viscosity~\cite{Ivanov:2010my}.

Regardless of the specific nature of the fluid anisotropy,
its presence may result in considerable changes in the characteristics of relativistic stars, as demonstrated, for
example, in Refs.~\cite{HH1975,anis_stars,Horvat:2010xf,Herrera:2013fja}.
But apart from that,
when considering magnetized matter, one might also expect the influence of the anisotropy on
the structure of the magnetic field.
The aim of the present paper is to get some insight into this question.

We emphasize that our goal
is not in itself to model more or less realistic distributions of the magnetic field.
We just want to demonstrate some
effects associated with the presence of the fluid anisotropy.
To do this, we employ a simplified model of a
poloidal magnetic field, i.e., the field lying in the meridional planes.
 Such a field is known to be unstable
\cite{Wright:1973,Markey:1973}, and only configurations consisting
of poloidal and toroidal fields with comparable strengths can be stable~\cite{Wright:1973,Braithwaite:2005md}. Nevertheless, the model
with a purely poloidal field, on the one hand, reflects the main qualitative characteristics of more realistic magnetic field structures,
and, on the other hand, within this model it is quite easy to carry out calculations and to keep track of changes coming from the fluid anisotropy.

The presence of such a dipole field will inevitably lead to deformations of the configuration, so that it is no longer spherically symmetric
but possesses instead axial symmetry.
However, since the energy density of the magnetic field discussed here
is much smaller than the energy density of the neutron matter, we follow Ref.~\cite{Konno:1999zv} and consider these
deformations as second-order corrections to the background spherically symmetric configuration. Using this fact, we
make a perturbative expansion of the field equations. In this case the equations for fluid perturbations are not independent
 but subject to an integrability condition, which in turn gives an equation for the current. Its solution  depends considerably
on the degree of fluid anisotropy, and this ultimately causes changes in the structure of the magnetic field of the
configurations under consideration.
Note that this approach was used by us earlier in Ref.~\cite{Aringazin:2014rva} to describe magnetic fields in mixed systems
consisting of a wormhole filled by a strongly magnetized isotropic or anisotropic neutron fluid.

The paper is organized as follows:
In Sec.~\ref{statem_prob} we present the statement of the problem
and derive the corresponding general-relativistic equations
for the systems under consideration.
In Sec.~\ref{num_calc} we numerically solve these equations
for isotropic and anisotropic cases.
Comparing the results, we demonstrate the influence
of the fluid anisotropy
on the structure and strength of the magnetic fields.
Finally, in Sec.~\ref{conclusion} we summarize the results obtained.

\section{Statement of the problem and equations}
\label{statem_prob}

Here we consider gravitating configurations consisting of a strongly magnetized anisotropic fluid.
Our purpose is to examine how the presence of the fluid anisotropy affects the structure of the magnetic field.
In doing so, we make use of the following simplifying assumptions \cite{Konno:1999zv}:

\begin{itemize}
  \item
 The magnetic field is modeled in the form of an
axisymmetric, poloidal magnetic field produced by toroidal electric currents.
Its presence in the system will in general result in a deviation of the shape of the configuration from spherical symmetry.
However, for values of the magnetic field strength of the order of
$10^{12}-10^{15}$~G, which will be discussed below,
deviations from the spherical shape are negligible,
since the energy of the magnetic field is much smaller
than the gravitational energy~\cite{Sotani:2006at}.
This permits us to neglect in the zeroth approximation the deformations of the configuration associated
with the magnetic field and to consider such deformations as a second-order effect.

\item We do not take into account rotational deformations and consider  only static equilibrium  configurations.

\item The neutron matter is modeled by the simplest polytropic equation of state.

\item The interior of a star is assumed to be
a perfectly conducting medium free of electric charges and fields.
\end{itemize}

Consistent with this,
in Sec.~\ref{backg_eqs}
we write down the background
equations for a spherically symmetric case, and
in Sec.~\ref{magn_field_eq} we derive an equation for the magnetic field.
Finally, in Sec.~\ref{integr_cond} we perturb the background solutions
and employ the obtained perturbed equations to derive an equation for the current.

\subsection{Background equations}
\label{backg_eqs}

To derive the Einstein equations and the Tolman-Oppenheimer-Volkoff equation for the fluid, we choose
the spherically symmetric line element in the form
\begin{equation}
\label{metric_schw}
ds^2=e^{\nu}(dx^0)^2-e^{\lambda}dr^2-r^2 \left(d\Theta^2+\sin^2\Theta\, d\phi^2\right),
\end{equation}
where $\nu$ and $\lambda$ are functions of the radial coordinate $r$ only,
and $x^0=c\, t$ is the time coordinate.
We refer to these equations as the {\sl background equations}.

As a matter source in these background equations, we take an anisotropic
fluid, i.e., the fluid for which
 the radial,~$p_r$, and tangential, $p_t$, components of pressure are not equal to each other.
For such a fluid,
the energy-momentum tensor can be taken in the form (see, e.g., Ref.~\cite{Herrera:2013fja})
\begin{equation}
\label{fluid_emt_anis}
T_{\nu (\text{fl})}^\mu=\left(\varepsilon +p_t\right)u^\mu u_\nu-\delta_\nu^\mu p_t+\left(p_r-p_t\right)s^\mu s_\nu,
\end{equation}
where $\varepsilon$ is the fluid energy density. The radial unit vector  $s^\mu$ is defined as
$s^\mu=\left(0, e^{-\lambda/2},0,0\right)$, with $s^\mu s_\mu=-1$ and $s^\mu u_\mu=0$. The energy-momentum tensor then contains the following
nonzero diagonal components: $T_{\nu (\text{fl})}^\mu=\left(\varepsilon,-p_r, -p_t, -p_t\right)$.

Using the metric \eqref{metric_schw} and the energy-momentum tensor \eqref{fluid_emt_anis}, one can obtain the $(^t_t)$ and $(^r_r)$ components of the Einstein
equations,
\begin{eqnarray}
\label{Einstein-00_anis}
&&-e^{-\lambda}\left(\frac{1}{r^2}-\frac{\lambda^\prime}{r}\right)+\frac{1}{r^2}
=\frac{8\pi G}{c^4} \varepsilon,
 \\
\label{Einstein-11_anis}
&&-e^{-\lambda}\left(\frac{1}{r^2}+\frac{\nu^\prime}{r}\right)+\frac{1}{r^2}
=-\frac{8\pi G}{c^4} p_r,
\end{eqnarray}
where $G$ is the gravitational constant
and the prime denotes differentiation with respect to $r$. In turn, the $\mu=r$ component
of the conservation law,
$T^\nu_{\mu (\text{fl}); \nu}=0$, yields the generalized Tolman-Oppenheimer-Volkoff equation for the anisotropic
matter:
\begin{equation}
\label{conserv_anis}
\frac{d p_r}{d r}=-G\left(\varepsilon+p_r\right)\frac{M+4\pi r^3 p_r/c^2}{r\left(c^2 r-2 G M\right)}+\frac{2}{r}\left(p_t-p_r\right).
\end{equation}
In obtaining this equation, we have used Eq.~\eqref{Einstein-11_anis} and introduced a new function $M(r)$, defined as
$$
e^{-\lambda}=1-\frac{2 G M(r)}{c^2 r}.
$$
Substituting this expression into Eq.~\eqref{Einstein-00_anis}, we also have
\begin{equation}
\label{mass_eq}
\frac{d M}{d r}=\frac{4\pi}{c^2} r^2 \varepsilon.
\end{equation}

For a complete description of the background configuration, Eqs.~\eqref{conserv_anis} and \eqref{mass_eq}
must be supplemented by an equation of state (EOS) for the fluid.
Here we restrict ourselves
to a simple barotropic EOS where the pressure is a function of the mass density $\rho$.
In this case, there are two possibilities of specifying the EOS. First, one may take two different EOSs
for the radial and the tangential  components of pressure, $p_r=p_r(\rho)$
and $p_t=p_t(\rho)$. Second, it is possible to restrict oneself to one EOS,
say, $p_r=p_r(\rho)$, but, in addition to this, one may then assign the function
$\Delta\equiv p_t-p_r$ which appears in Eq.~\eqref{conserv_anis}.
This function is called the anisotropy factor~\cite{Herrera:1985}.

Here we employ the second possibility, using for the radial pressure the simplest polytropic EOS in the form
\begin{equation}
\label{eos_anis}
p_r=K \rho^\gamma, \quad \varepsilon =\rho c^2,
\end{equation}
where $K, \gamma$ are constants whose value depends on the properties of the fluid under
consideration. Such an EOS was used in Ref.~\cite{Tooper:1964} in modeling general relativistic
isotropic fluid spheres and in Refs.~\cite{Horvat:2010xf,Herrera:2013fja}
in describing objects with an anisotropic pressure.

Following Ref.~\cite{Horvat:2010xf}, here we take
the anisotropy factor $\Delta$ in the following form:
\begin{equation}
\label{anis_fact}
\Delta\equiv p_t-p_r=\alpha p_r \mu,
\end{equation}
where $\alpha$ is the anisotropy parameter, and the function $$\mu=\frac{2 G M(r)}{c^2 r}$$ is called the compactness.

The choice \eqref{anis_fact} has the following attractive features~\cite{Horvat:2010xf}: (i) since at $r\to 0$ the compactness $\mu \sim r^2$,
 the anisotropy factor vanishes at the center (where the fluid becomes isotropic),
and this ensures the regularity of the right-hand side
of Eq.~\eqref{conserv_anis} (for other possibilities to obtain regular solutions without imposing the requirement for the anisotropy to vanish
at the center,
see Ref.~\cite{HH1975});
(ii) the anisotropy factor thus defined makes itself felt only for what are essentially relativistic configurations, for which
$\mu\sim {\cal O}(1)$. This agrees with generally accepted folklore, according to which the fluid anisotropy may play an important role
only at high densities of matter~\cite{HH1975,anis_stars,Horvat:2010xf}.

Note that in the absence of $\mu$ on the right-hand side of  \eqref{anis_fact}
the ratio of pressures $p_t/p_r$
would be constant along the radius, and the problem would become analogous to the case of
Refs.~\cite{HH1975,Aringazin:2014rva}.
However, such an ansatz for the fluid anisotropy seems to be less realistic than
that of~\eqref{anis_fact}. Another ansatz for the anisotropy can be found in Refs.~\cite{HH1975,anis_stars}.

\subsection{Equation for the magnetic field}
\label{magn_field_eq}

As an ansatz for the magnetic field, choose an
axisymmetric, poloidal magnetic field, which is created by a 4-current
$j_{\mu}=(0,0,0,j_{\phi})$~\cite{Konno:1999zv}. For such a current,
the electromagnetic 4-potential $A_{\mu}$
has only a $\phi$-component $A_{\mu}=(0,0,0,A_{\phi})$. In this case, taking into account the nonvanishing components of
the electromagnetic field tensor $F_{r \phi}=\partial A_{\phi}/\partial r$ and $F_{\Theta \phi}=\partial A_{\phi}/\partial \Theta$,
the Maxwell equations give in the metric \eqref{metric_schw}
the following elliptic equation:
\begin{equation}
\label{maxw_A}
e^{-\lambda}
\frac{\partial^2 A_{\phi}}{\partial r^2}+\frac{1}{2} \left(\nu^\prime-\lambda^\prime\right)e^{-\lambda}\frac{\partial A_{\phi}}{\partial r}+
\frac{1}{r^2}\frac{\partial^2 A_{\phi}}{\partial \Theta^2}-
\frac{1}{r^2}\cot \Theta\frac{\partial A_{\phi}}{\partial \Theta}=-\frac{1}{c} j_{\phi}.
\end{equation}
The solution of this equation is sought as an expansion of
 the potential $A_{\phi}$ and the current $j_{\phi}$ as follows~\cite{Regge:1957,Konno:1999zv}:
\begin{eqnarray}
\label{expan_A}
&& A_{\phi}=\sum_{l=1}^\infty a_l(r)\sin \Theta \frac{d P_l(\cos \Theta)}{d\Theta},\\
&& j_{\phi}=\sum_{l=1}^\infty j_l(r)\sin \Theta \frac{d P_l(\cos \Theta)}{d\Theta},
\end{eqnarray}
where $P_l$ is the Legendre polynomial of degree $l$. Substituting these expressions into Eq.~\eqref{maxw_A}, we have
\begin{equation}
\label{maxw_A_expan}
e^{-\lambda}
a_l^{\prime \prime}+\frac{1}{2}\left(\nu^\prime-\lambda^\prime\right)e^{-\lambda} a_l^\prime-\frac{l(l+1)}{r^2}a_l=-\frac{1}{c}j_l.
\end{equation}
The solution of this equation is sought for a given current~$j_l$,
an equation for which will be derived in the next subsection. Since here we consider only a dipole magnetic field
for which $l=1$, for convenience, we drop the subscript~1 on $a$ and~$j$ hereafter.

\subsection{Integrability condition}
\label{integr_cond}

For the magnetic field under consideration,
the current $j$ is not arbitrary but must satisfy an integrability condition~\cite{Bocquet:1995je,Konno:1999zv}.
To obtain this condition, we make use of the fact that the magnetic field induces only small deviations
in the shape of the background spherically symmetric configuration.
To describe  these deviations, let us employ the approach adopted in Ref.~\cite{Konno:1999zv} and expand
the metric in multipoles around the spherically symmetric
spacetime. Then the deformations of the metric and the fluid are regarded as second-order perturbations, whereas the electromagnetic field potential and the
current are regarded as first-order perturbations. In this case the corresponding metric can be taken in the form
\begin{eqnarray}
\label{pert_metr}
&& ds^2  =   e^{\nu (r)} \left\{ 1 + 2 \left[ h_{0}(r) + h_{2}(r)
          P_{2}( \cos \Theta ) \right] \right\} (dx^0)^2  
   -  e^{\lambda(r)}\left\{ 1 + \frac{2 e^{\lambda(r)}}{r}
          \left[ m_{0}(r) + m_{2}(r) P_{2}( \cos \Theta )
          \right] \right\} dr^2
          \nonumber \\
  & & -  r^2 \left[ 1 + 2 k_{2}(r) P_{2}( \cos \Theta )
        \right] \left( d \Theta^2
        +  \sin^2 \Theta d \phi^2 \right) \!,
\end{eqnarray}
where $h_0$, $h_2$, $m_0$, $m_2$, and $k_2$ are the second-order corrections of the metric, and $P_2$ denotes
the Legendre polynomial of order~2.

The total energy-momentum tensor for the system under consideration is
\begin{equation}
\label{EMT_total}
T_{\nu }^\mu=\left(\varepsilon +p_t\right)u^\mu u_\nu-\delta_\nu^\mu p_t+\left(p_r-p_t\right)s^\mu s_\nu
-F^\mu_\alpha F_\nu^\alpha+\frac{1}{4}\delta^\mu_\nu F_{\alpha \beta}F^{\alpha \beta}.
\end{equation}
Using the relation between the pressures \eqref{anis_fact}, we can eliminate the tangential pressure; i.e.,
henceforth we will work only with the radial pressure.

Expand now the fluid energy density and the pressure in the form
\begin{eqnarray}
\label{expans}
&&\varepsilon(r,\Theta)=\varepsilon_0+\frac{\varepsilon_0^\prime}{p_0^\prime}\left(\delta p_0+\delta p_2 P_2\right),\\
&&p(r,\Theta)=p_0+\delta p_0+\delta p_2 P_2,
\end{eqnarray}
where the background solutions $\varepsilon_0, p_0$ and the perturbations $\delta p_0, \delta p_2$
depend on $r$ only.
(For convenience, we hereafter drop the subscript ``$r$'' on $p$.)
Substituting   these expressions into the conservation law, $T^\mu_{\nu; \mu}=0$,
and using the metric \eqref{pert_metr},
we  obtain the following $\nu=r$ and $\nu=\Theta$ components:
\begin{eqnarray}
\label{p2expres1}
&&\delta p_2^\prime=2\alpha \mu p_0 k_2^\prime-\left(\varepsilon_0+ p_0\right)h_2^\prime-\frac{1}{2}\left[\nu^\prime\left(1+\frac{\varepsilon_0^\prime}{p_0^\prime}
\right)-\frac{4\alpha \mu}{r}\right]\delta p_2
-\frac{2}{3}\frac{a^\prime}{r^2}\frac{j}{c},\\
\label{p2expres2}
&&(1+\alpha\mu)\delta p_2=-\left[(1+\alpha\mu)p_0+\varepsilon_0\right] h_2-\frac{\alpha\mu}{1-\mu}\frac{m_2 p_0}{r}
-\frac{2}{3}\frac{a}{r^2}\frac{j}{c}.
\end{eqnarray}

The function $\nu^\prime$ entering into Eq.~\eqref{p2expres1} can be found from the background equations in the form
\begin{equation}
\label{nuprime}
\nu^\prime=\frac{2 p_0^\prime}{\varepsilon_0+ p_0}\left(\frac{2\alpha\mu}{r}\frac{p_0}{p_0^\prime}-1\right).
\end{equation}

The integrability condition for Eqs.~\eqref{p2expres1} and \eqref{p2expres2} gives the following equation for the current:
\begin{equation}
\label{ic_current}
j^\prime+F j+N=0
\end{equation}
 with
\begin{equation}
\label{funF}
 F=-\frac{1}{1+\alpha\mu}\left\{\frac{2}{r}-\frac{1}{2}\nu^\prime\left(1+\frac{\varepsilon_0^\prime}{p_0^\prime}\right)+
\alpha^2\mu^2\left(\frac{2}{r}+\frac{a^\prime}{a}\right)+
\alpha\mu\left[\frac{4}{r}+\frac{\mu^\prime}{\mu}+\frac{a^\prime}{a}-\frac{1}{2}\nu^\prime\left(1+\frac{\varepsilon_0^\prime}{p_0^\prime}\right)\right]
\right\}
\end{equation}
and
\begin{eqnarray}
\label{funN}
N=&&\alpha\frac{3 c \mu r^2}{2 a}\Big\{\frac{8\pi G}{3 c^4}(1-\mu)a^\prime\left[
 a^\prime p_0^\prime+p_0\left(
\left[\mu^\prime\frac{1-\mu(2+\alpha\mu)}{\mu(1-\mu)(1+\alpha\mu)}+\frac{1}{2}\nu^\prime\left(1+\frac{\varepsilon_0^\prime}{p_0^\prime}\right)
-\frac{2\alpha\mu}{r}\right]a^\prime+2a^{\prime\prime}
\right)
\right]
\nonumber\\
&&-\left(\varepsilon_0+p_0\right)h_2^\prime+2(1+\alpha\mu)p_0 k_2^\prime-\left[
\frac{\varepsilon_0+p_0}{1+\alpha\mu}\frac{\mu^\prime}{\mu}+\frac{2}{r}\left(\varepsilon_0-p_0\frac{\varepsilon_0^\prime}{p_0^\prime}\right)
\right]h_2
\Big\}.
\end{eqnarray}
Note that all terms in Eq.~\eqref{ic_current} are of first order.
Indeed, $j$ itself is of first order,
and the function $F$ is of zeroth order.
Finally, the function $N$ is also of first order,
since the potential $a$,
which is of first order, enters in the denominator,
and the numerator contains the second-order terms.

In the anisotropic case ($\alpha\neq 0$), the function $N$
is nonzero. To evaluate it, one needs to find
the functions $h_2$ and $k_2$ appearing in \eqref{funN}, which are determined by the following set of equations
deduced from the Einstein equations:
\begin{eqnarray}
\label{pert_Einst_G11_l2}
&&h_2^\prime+\left(1+\frac{r}{2}\nu^\prime\right)k_2^\prime=\frac{1}{1-\mu}\left\{
\frac{3}{r} h_2+\frac{2}{r} k_2+\frac{1+r\nu^\prime}{r^2}m_2
-\frac{4\pi G}{3 c^4}\frac{1}{r}\left[
(1-\mu)a^{\prime 2}+4\frac{a^2}{r^2}\right]\right\},\\
\label{pert_Einst_G21_l2}
&&h_2^\prime+k_2^\prime=\left(\frac{1}{r}-\frac{1}{2}\nu^\prime\right)h_2+\left(\frac{1}{r}+\frac{1}{2}\nu^\prime\right)\frac{m_2}{r(1-\mu)}+
\frac{16\pi G}{3 c^4}\frac{a a^\prime}{r^2}, \\
\label{pert_Einst_G3_G22}
&& h_2+\frac{m_2}{r(1-\mu)}=\frac{8\pi G}{3 c^4}(1-\mu)a^{\prime 2},\\
\label{pert_Einst_G00_l2}
&& k_2^{\prime\prime}-\left[\frac{\mu^\prime}{2(1-\mu)}-\frac{3}{r}\right]k_2^\prime
-\frac{1}{r^2(1-\mu)}\left[2 k_2+m_2^\prime+\frac{3 m_2}{r(1-\mu)}\right]
=\frac{4\pi G}{c^4}\frac{1}{1-\mu}\left(-\frac{\varepsilon_0^\prime}{p_0^\prime}\delta p_2
+\frac{1-\mu}{3}\frac{a^{\prime 2}}{r^2}-\frac{4}{3}\frac{a^2}{r^4}\right).
\end{eqnarray}

Notice here that in deriving the expression \eqref{funN},
the function $m_2$, which appears in \eqref{p2expres2}, has been eliminated
by using Eq.~\eqref{pert_Einst_G3_G22}.

For the isotropic case ($\alpha=0$), Eq.~\eqref{ic_current} can be integrated analytically to give
\begin{equation}
\label{ic_current_isot}
j=c_0 r^2 (\varepsilon_0+p_0),
\end{equation}
where $c_0$ is an integration constant. This expression
coincides with the one obtained in Ref.~\cite{Konno:1999zv} [see their Eq.~(25)].

It should be emphasized that the small deviations in the shape of the configuration
described by the second-order corrections to the metric $h_2$, $k_2$, and $m_2$ affect the current, and correspondingly the structure
of the magnetic field, only if the fluid anisotropy is present.
The question of how this manifests itself will be discussed in the next section.

\section{Numerical results}
\label{num_calc}

In this section we numerically integrate the obtained equations for isotropic ($\alpha=0$)
and anisotropic ($\alpha\neq 0$) cases.
Our goal is to exhibit differences in distribution of the magnetic field for these two cases.

\subsection{Background equations}

For numerical calculations,
it is convenient to rewrite Eqs.~\eqref{conserv_anis} and \eqref{mass_eq} in terms of dimensionless variables.
To do this, we employ the usual reparametrization of the fluid density~\cite{Zeld},
\begin{equation}
\label{theta_def}
\rho=\rho_{c} \theta^n~,
\end{equation}
where $\rho_{c}$ is the density of the  fluid at the
center of the configuration, and the constant $n$, called the polytropic index,
is related to $\gamma$ from the EOS \eqref{eos_anis} via $n=1/(\gamma-1)$.
Introducing the dimensionless variables
\begin{equation}
\label{dimless_var_NS}
\xi=\frac{r}{L}, \quad v(\xi)=\frac{M(r)}{4\pi \rho_c L^3},
\quad \text{where} \quad L=\sqrt{\frac{\sigma(n+1)c^2}{4\pi G \rho_c}},
\end{equation}
we have from Eqs.~\eqref{conserv_anis} and \eqref{mass_eq}
\begin{eqnarray}
\label{conserv_anis_dmls_NS}
&&\sigma(n+1)\frac{d\theta}{d\xi}=-\sigma(n+1)(1+\sigma\theta)\frac{v+\sigma\xi^3\theta^{n+1}}{\xi^2\left(1-\mu\right)}+
2\alpha\sigma\frac{\mu\, \theta}{\xi},
\\
\label{mass_eq_dmls_NS}
&&\frac{d v}{d\xi}=\xi^2 \theta^n.
\end{eqnarray}
Here $\sigma=K \rho_{c}^{1/n}/c^2=p_c/(\rho_{c} c^2)$ is a constant
related to the pressure $p_c$ of the fluid at the center, 
and ${\mu=2\sigma(n+1)v/\xi}$. The parameter $\sigma$, also called the relativity parameter \cite{Tooper:1964},
measures the deviation of a configuration from nonrelativistic systems for which $\sigma \ll 1$.

These equations are to be solved for given $n$ and $K$
subject to the boundary conditions given in the neighborhood of the center
by the following expansions:
\begin{equation}
\label{bound}
\theta\approx 1+\frac{1}{2}\theta_2\xi^2, \quad v\approx \frac{1}{3}\xi^3,
\quad \text{where} \quad \theta_2=-(\sigma+1)(\sigma+1/3)+\frac{4}{3}\alpha\sigma.
\end{equation}

Notice that in the case of an isotropic fluid ($\alpha=0$), $\theta_2$ is always negative, and correspondingly
the fluid has a maximum density at the center. For the anisotropic fluid  ($\alpha\neq 0$), a situation may occur
where $\theta_2>0$; i.e., the fluid will have a local minimum of the density at the center, and a maximum will be located somewhere
between the center and the edge of the star.
An example of such a configuration can be found in Ref.~\cite{Horvat:2010xf}.

Using these boundary conditions, we numerically solved Eqs.~\eqref{conserv_anis_dmls_NS} and \eqref{mass_eq_dmls_NS}, started the
solutions near the origin (i.e., near $\xi \approx 0$), and solved out to
a point $\xi=\xi_b$, where the function $\theta$ became zero (the boundary of the fluid where $p(\xi_b)=0$).

Note here that the  radial coordinate $r$  from the metric \eqref{metric_schw}
describes the areal radius of a sphere with area $4\pi r^2$.
Another physically relevant radial coordinate is given by the coordinate $\tilde{\xi}$  associated with the proper radius, which
represents the true distance from the center. It is defined as follows:
\begin{equation}
\label{prop_coord}
\tilde{\xi}=\int_0^\xi e^{\lambda/2}d \xi=\int_0^\xi \frac{d \xi}{\sqrt{1-\mu}}.
\end{equation}
Then, the proper radius of the fluid $R$ is obtained in dimensional variables as $R=\tilde{\xi}_b L$.

\subsection{Magnetic field equations}

We rewrite now the perturbative equations ~\eqref{maxw_A_expan} and \eqref{ic_current}
for the electromagnetic potential and the current through
 appropriate dimensionless variables:
\begin{equation}
\label{dimless_xi_v2}
\bar{a}(\xi)=\frac{8\pi G }{c^3}\sqrt{\frac{\rho_c}{2\sigma (n+1)}}\,a(r),
\quad \bar{j}(\xi)=\sqrt{\frac{2\sigma (n+1)}{\rho_c c^4}}\, j(r).
\end{equation}
Taking also into account  \eqref{dimless_var_NS}, we have
\begin{eqnarray}
\label{Maxw_dmls}
&&(1-\mu)\left[
\bar{a}^{\prime \prime}+\frac{1}{2}\left(\nu^\prime -\frac{\mu^\prime}{1-\mu}\right)\bar{a}^\prime
\right]-\frac{2}{\xi^2}\bar{a}=-\bar{j}, \\
\label{current_anis_dmls}
&&\bar{j}^\prime+\bar{F} \bar{j}+\bar{N}=0,
\end{eqnarray}
where
$\bar{F}, \bar{N}$ are the dimensionless expressions
obtained from $F$ and $N$, Eqs.~\eqref{funF} and \eqref{funN}.

The boundary conditions for these equations are given
by an expansion in the neighborhood of the center,
\begin{equation}
\label{bound_cond_all2}
\bar{a} \approx \frac{1}{2}\, a_c \xi^2,
\quad \bar{j} \approx \frac{1}{2} \, j_c \xi^2,
\end{equation}
where  $a_c, j_c$ are free parameters.

In the isotropic case $\bar N=0$,
and Eq.~\eqref{current_anis_dmls} has the solution in the form of \eqref{ic_current_isot}.

When an anisotropy is present,
Eqs.~(\ref{Maxw_dmls}) and~\eqref{current_anis_dmls} are solved together with
the Einstein equations~\eqref{pert_Einst_G11_l2}-\eqref{pert_Einst_G00_l2}
for the second-order perturbations.
To obtain regular solutions,
we choose the boundary conditions for them in the form
\begin{equation}
\label{bound_cond_pert}
h_2 \approx \frac{1}{2}h_{2c}\xi^2,
\quad k_2 \approx \frac{1}{2}k_{2c}\xi^2.
\end{equation}
In turn,
the boundary condition for the function $m_2$ can be found from Eq.~\eqref{pert_Einst_G3_G22}.
Using these boundary conditions in Eq.~\eqref{pert_Einst_G21_l2} or in Eq.~\eqref{pert_Einst_G00_l2},
one can find the following constraint on the expansion parameters:
$$
h_{2c}+k_{2c}=\frac{2}{3}a_c^2.
$$

Thus, we have three free parameters,
 $a_c, j_c$, and  $h_{2c}$ or $k_{2c}$,
which are chosen in such a way as to (i)~get a required value of the surface magnetic field,
(ii) provide regularity of the perturbed solutions at the center,
$\xi=0$, and (iii)~obtain asymptotically decaying solutions for
$\xi \rightarrow \infty$.

In particular, it is necessary to derive the known external
solution~\cite{Konno:1999zv}, according to which beyond the fluid
the electromagnetic field potential decays as
$$
\bar{a}\sim -\xi^2\left[\ln{(1-\mu)}+\mu+\frac{1}{2}\mu^2\right].
$$
In this solution,  $\mu$ corresponds to the external vacuum solution for the background configuration with the mass concentrated
inside the radius $\xi_b$.

Finally, by solving the equation for the electromagnetic field potential, we can find
the strength of the magnetic field,
which is given by the following tetrad components  (i.e., the components
measured by a locally inertial observer):
\begin{equation}
\label{streng_magn_dmls_NS}
B_{\hat{r}}=-F_{\hat{\Theta} \hat{\phi}}= c\sqrt{\frac{2\rho_{c}}{\sigma(n+1)}}\frac{\bar{a}}{\xi^2}\cos{\Theta}, \quad
B_{\hat{\Theta}}=F_{\hat{r} \hat{\phi}}=- c\sqrt{\frac{\rho_{c}}{2\sigma(n+1)}}\frac{\sqrt{1-\mu}}{\xi}\,\bar{a}^\prime\sin{\Theta}.
\end{equation}

\subsection{Structure of the magnetic field}
\label{mag_f_str}

Consider now
magnetic field distributions for the configurations with different values of the anisotropy parameter~$\alpha$.
In Ref.~\cite{Horvat:2010xf}, finite-size, regular solutions for compact relativistic configurations with the EOS~\eqref{eos_anis}
 and $-2 \leq \alpha \leq 4$ have been found. For such systems,
the mass-radius curve has a turning point corresponding to the maximum allowed mass at which the system is still stable.
It was shown that as $\alpha$ decreases, the maximum mass decreases as well.
In particular, when
the polytropic index $n=1$ ($\gamma=2$) and $K=6.67\times 10^4\, \text{cm}^{5}\text{g}^{-1}\text{s}^{-2}$ (in our units),
the maximum mass  $M_{\text{max}}\approx 1 M_\odot$ for the system with $\alpha=-2$ and $M_{\text{max}}\approx 2.14 M_\odot$ for the case of  $\alpha=4$.

\begin{figure}[p]
\begin{minipage}{1\linewidth}
  \begin{center}
  \includegraphics[width=12cm]{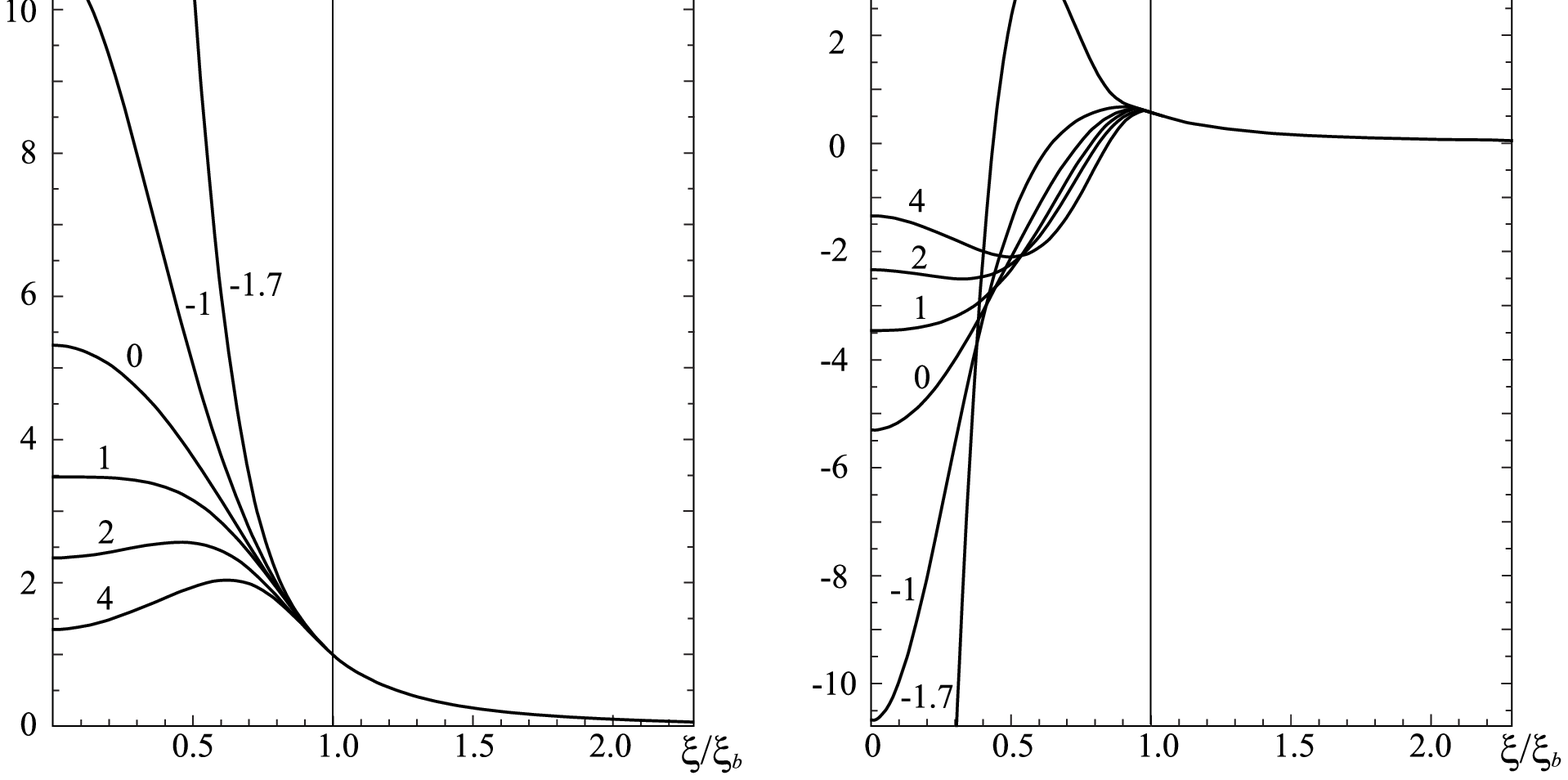}
\vspace{.3cm}
  \end{center}
\end{minipage}\hfill
\begin{minipage}{1\linewidth}
  \begin{center}
  \includegraphics[width=7.cm]{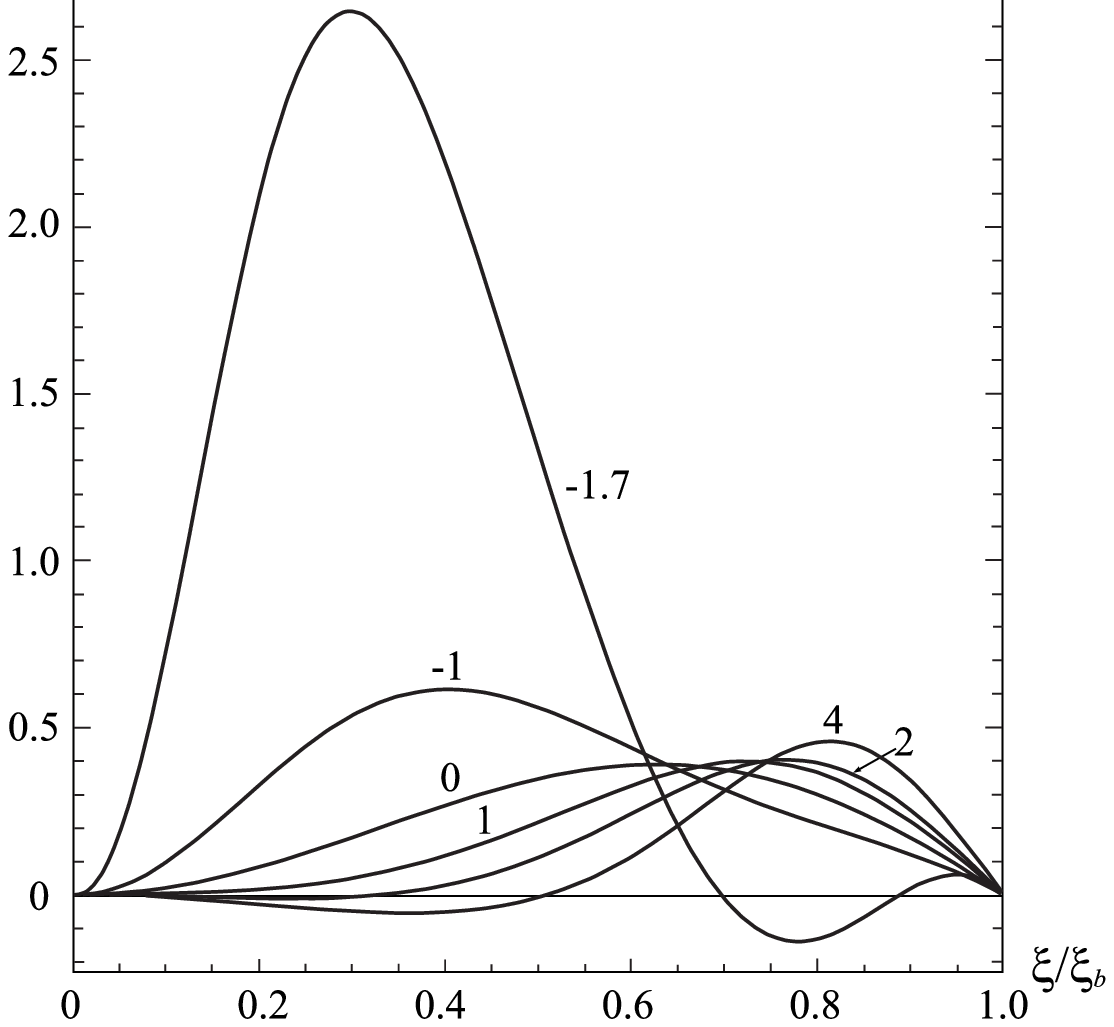}
  \end{center}
\end{minipage}\hfill
  \caption{The tetrad components $B_{\hat{r}}$ and $B_{\hat{\Theta}}$
of the magnetic field (in units of the surface strength
of the magnetic field $B_s$ at the pole) evaluated
on the symmetry axis ($\Theta=0$) and
in the equatorial plane ($\Theta=\pi/2$), respectively,
and the dimensionless current $\bar{j}$
are shown as functions of the relative radius $\xi/\xi_b$.
The numbers near the curves denote the values of the anisotropy parameter $\alpha$.
The thin vertical lines correspond to the boundary of the fluid.
All configurations have the same baryon mass,
$M_b \approx 1.23 M_\odot$ (for other characteristics, see Table~\ref{tab1}).
For  $\alpha=-1.7$, the central values of $B_{\hat{r}}/B_s$ and $B_{\hat{\Theta}}/B_s$ are of the order of $44$ (modulus).
}
\label{MF_NS_fig}
\vspace{.3cm}
\centering
  \includegraphics[height=6cm]{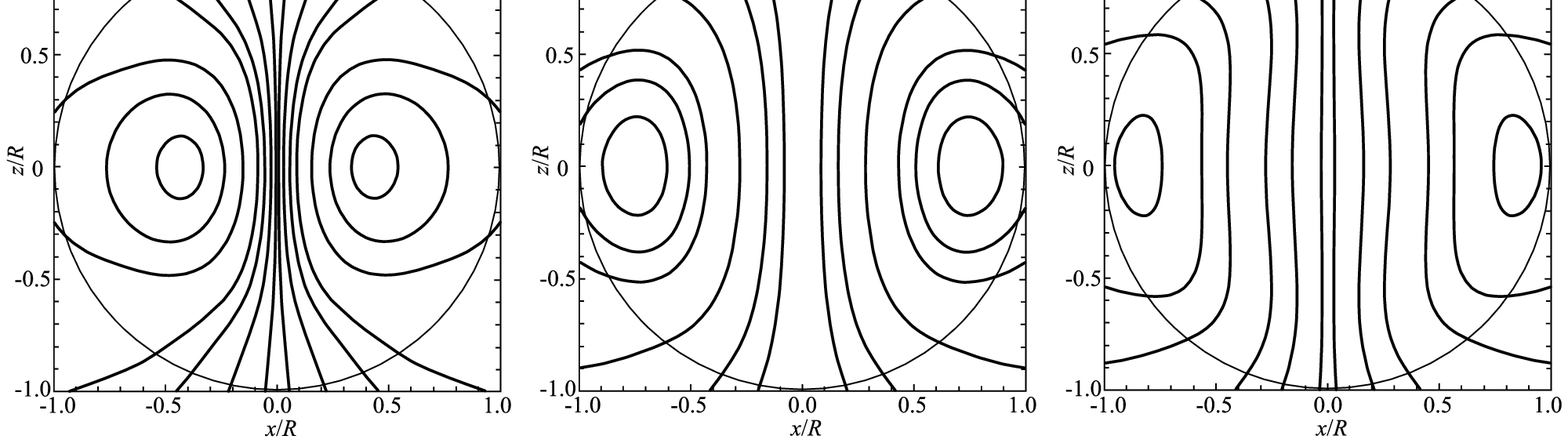}
\caption{Magnetic field lines for the configurations of Fig.~\ref{MF_NS_fig}.
The plots are made in a meridional plane $\phi=\text{const.}$ spanned by the coordinates
$x=r \sin{\Theta}$ and $z=r \cos{\Theta}$.
The circles denote the boundary of the neutron fluid,
possessing the radius $R$.
}
\label{fig_magn_field_lines_NS}
\end{figure}

\begin{figure}[h!]
\centering
  \includegraphics[height=7cm]{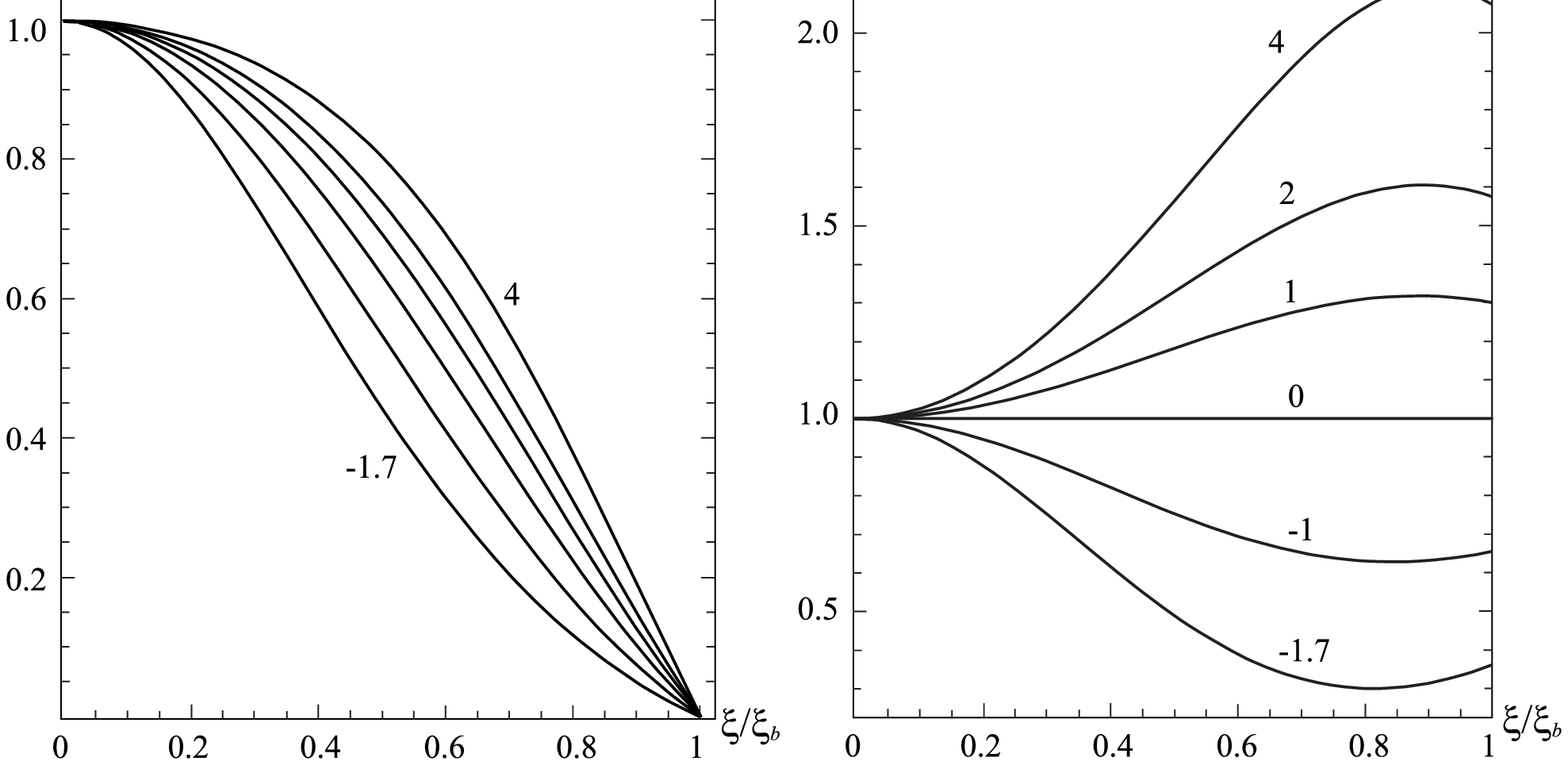}
\caption{Energy density $T_{t (\text{fl})}^t=\varepsilon$ of the neutron fluid
(in units of $\rho_{c} c^2$) and the ratio $p_t/p_r$ of the tangential pressure to the radial pressure
for the configurations of Fig.~\ref{MF_NS_fig}.
For the left panel, the values of the anisotropy parameter $\alpha$
are successively decreased from $\alpha=4$
(for the top curve) to $\alpha=-1.7$ (for the bottom curve).
}
\label{NS_energ_dens_fig}
\end{figure}

Here we use the configurations of Ref.~\cite{Horvat:2010xf} as background systems in which the magnetic field is added.
In order to show explicitly the influence of the fluid anisotropy on the structure and strength of the magnetic field inside the star,
we will proceed in the following way:

(i) Following Ref.~\cite{Bocquet:1995je}, we fix the number of baryons contained in each of the configurations under consideration
and compare the configuration supported by an isotropic fluid with the anisotropic systems.
This approach implies that systems with the same baryon number (or equivalently, the same baryon mass $M_b$) may have different physical
characteristics (total masses, sizes, an internal structure, etc.), depending both on the type of EOS and the presence of ultra-strong magnetic fields~\cite{Bocquet:1995je}
and on the presence of fluid anisotropy.
As will be shown below, the presence of the latter will in turn lead to considerable changes in the structure of the magnetic field.

(ii) All the systems in question are modeled by the EOS~\eqref{eos_anis} with some fixed
polytropic parameters. This allows us to keep track of changes in the structure of the magnetic field coming from just the fluid anisotropy,
but not those which depend on the form of EOS.

(iii) Depending on the value of the anisotropy parameter $\alpha$, the total mass $M$ of the configurations will change.
(Note that the total mass is understood here to be only the mass of the background configuration
without taking account of the mass coming from the magnetic field that is negligibly small compared with the mass of the fluid.)
In order to obtain the required fixed baryon mass $M_b$ for the systems with different $\alpha$,
we will appropriately change $\rho_c$. In turn, the free parameters of the magnetic field
$a_c, j_c$ will be chosen  in such a way as to provide the required surface magnetic field $B_s$ (say, at the pole) for all configurations,
and also to obtain asymptotically vanishing solutions for the electromagnetic potential.

Proceeding in this manner, we obtained the results
shown in Figs.~\ref{MF_NS_fig} and \ref{fig_magn_field_lines_NS}.
All calculations were carried out using
the polytropic parameters $n=1$ ($\gamma=2$) and $K=6.67\times 10^4\, \text{cm}^{5}\text{g}^{-1}\text{s}^{-2}$.
Since for these parameters the system with $\alpha=-2$ possesses
the lowest maximum total mass $M_{\text{max}}\approx 1 M_\odot$, we took it as the reference mass.
Then, we calculated the baryon mass $M_b$ of this system, fixed this value and then calculated the corresponding total masses
for the systems with different $\alpha>-2$ but all having this fixed value of $M_b$.
The results of the calculations are listed in Table~\ref{tab1}.
Note that all the configurations thus obtained are energetically stable (cf.  Ref.~\cite{Horvat:2010xf}).

\begin{table}
\caption{Characteristics of the configurations exhibited in Fig.~\ref{MF_NS_fig}.
Here the central mass density $\rho_{c}$ of the neutron fluid
(in units of $10^{14} \text{g cm}^{-3}$),
the relativity parameter $\sigma$,
the total mass $M$ (in solar mass units),
and the proper radius of the fluid $R$, as given by Eq.~\eqref{prop_coord}, (in kilometers) are shown.
}
\vspace{.3cm}
\begin{tabular}{p{1.2cm}p{1.2cm}p{1.2cm}p{1.2cm}p{1.2cm}}
\hline \\[-5pt]
$\alpha $ & $\rho_{c}$ & $\sigma$  & $ M/M_\odot$   & $ R$ \\[2pt]
\hline \\[-7pt]
\end{tabular}\\
\begin{tabular}{.....}
-1.7 &48.00&	0.36&	1.02 & 9.44\\
-1.0 &27.60& 0.20&	1.05 & 10.29\\
0.0 &17.50&	0.13&	1.07 & 11.12\\
1.0 &12.78&	0.10&	1.08 & 11.77 \\
2.0 &10.02&	0.07&	1.09 & 12.31\\
4.0 &6.92&	0.05&	1.10 & 13.22\\
\hline
\end{tabular}
\label{tab1}
\end{table}

In Fig.~\ref{MF_NS_fig} the distribution of the component
$B_{\hat{r}}$ on the symmetry axis ($\Theta=0$),
the distribution of the component $B_{\hat{\Theta}}$
in the equatorial plane ($\Theta=\pi/2$),
and the current $\bar j$ are shown. It is seen that the magnitude and the distribution of the internal magnetic field
change considerably for the systems with different $\alpha$. In the case of
$\alpha>0$ ($\alpha<0$), the value of the component $B_{\hat{r}}$ is always less (greater) than that of the isotropic case.
In turn, the magnitude of the component
 $B_{\hat{\Theta}}$ for the anisotropic fluid may be either greater or less (modulus) than the one for the isotropic case,
 depending on the point where a comparison is performed.
It is interesting to note that for positive values of $\alpha$ the profiles of both components $B_{\hat{r}}$ and $B_{\hat{\Theta}}$ are flattened
over the radius
up to the point $\sim R/2$. On the other hand, for large negative values of $\alpha$,
there is a rapid growth of the central values of
$B_{\hat{r}}$ and $B_{\hat{\Theta}}$, which, for $\alpha = -1.7$, differ in order of magnitude from those of the isotropic case.

The distribution of the current also depends considerably on the value of $\alpha$, see Fig.~\ref{MF_NS_fig}.
For example, if in the isotropic fluid
the current always has a fixed sign
[cf. Eq.~\eqref{ic_current_isot}], then in the anisotropic case it may have variable sign  both for positive and for negative
$\alpha$.

The magnetic field lines are shown in Fig.~\ref{fig_magn_field_lines_NS},
where the spatial coordinates are given in units of the radius
of the fluid, $R$. The behavior of the field lines is qualitatively similar for different values of $\alpha$.

Note that the perturbed equations \eqref{maxw_A_expan}, \eqref{ic_current}-\eqref{pert_Einst_G00_l2} are invariant under
the transformation
$a, j \to \beta a, \beta j$ and $h_2, k_2, m_2 \to \beta^2 h_2, \beta^2 k_2, \beta^2 m_2$,  $\beta$ being any constant.
Correspondingly, the components of
the strength of the magnetic field given by Eq.~\eqref{streng_magn_dmls_NS} transform as
 $B_{\hat{r}}, B_{\hat{\Theta}}\to \beta B_{\hat{r}}, \beta B_{\hat{\Theta}}$. Then, if one simultaneously replaces
 $B_s$ by $\beta B_s$, the plots shown in Fig.~\ref{MF_NS_fig} are unchanged for any value of the field, and the
dimensional values (in gauss) of the strength of the magnetic field are obtained on multiplying these plots
by the required surface value $B_s$.
Of course, this holds true only in the approximation used here when one can neglect the influence of the magnetic field on the background solutions.
In particular, the obtained plots are applicable both to the ``classical pulsars''
(for which $B_s\sim 10^{12}$~G) and to
magnetars (for which $B_s\sim 10^{15}$~G).

It may also be noted that for the systems with negative
$\alpha$, the relativity parameter $\sigma$ is greater than that for the case of positive  $\alpha$ (see Table~\ref{tab1}).
That is, the smaller $\alpha$, the more relativistic matter is necessary
in order to have equilibrium configurations with a fixed number of baryons of the type we discuss here.
In turn, the growth of the relativity parameter $\sigma$  is accompanied by the following effects: (i)~a greater concentration of matter toward the center,
as demonstrated in the left panel of Fig.~\ref{NS_energ_dens_fig} where the distributions of the fluid energy density $T_{t (\text{fl})}^t$
are shown;
(ii)~for strongly relativistic systems with large negative $\alpha$ there exists the above-mentioned rapid growth of the strength of the magnetic
field  in the internal regions of the configurations, see Fig.~\ref{MF_NS_fig};
(iii)~for negative $\alpha$, a more rapid increase in the difference between the tangential and
radial pressures takes place
(as one can see, e.g., from a comparison of the curves for $\alpha=1$ and $\alpha=-1$ shown in the right panel of Fig.~\ref{NS_energ_dens_fig}).

\section{Conclusion}
\label{conclusion}

We have studied equilibrium, gravitating configurations consisting of a strongly magnetized fluid
with an anisotropic pressure. The fluid is described by the simplest
polytropic EOS \eqref{eos_anis}, and the anisotropy is modeled by the expression~\eqref{anis_fact}, which takes into account both
the local properties of the matter (through the pressure) and  the  quasilocal properties of the configuration (through the compactness).
Our goal was to clarify  the question of how the presence of such an anisotropy influences
the structure of the magnetic field, which
was modeled here in the form of a poloidal field  produced by toroidal electric currents.

For this purpose, we compared configurations with the same number of baryons but with different degrees of fluid anisotropy.
Having fixed the polytropic parameters, we kept track of changes in the internal structure of the magnetic field
coming from just the fluid anisotropy (for details, see Sec.~\ref{mag_f_str}). These changes
were then revealed by comparing the configurations with isotropic and anisotropic fluids.

In summary, the main results of the studies are as follows:

\begin{enumerate}
\itemsep=-0.2pt
\item[(i)] 
For an anisotropic fluid, an equation for the current is modified in such a way that it begins to ``feel'' the presence of small deformations of a star
associated with the presence of a dipole magnetic field [see Eq.~\eqref{ic_current}]. This happens both in the case of the quasilocal anisotropy considered here
and in the case where an anisotropy is constant along the radius (cf. Ref.~\cite{Aringazin:2014rva}).
In contrast to this, for configurations
supported by an isotropic fluid, the deformations due to the presence of an
axisymmetric magnetic field do not affect the structure of the magnetic field.

\item[(ii)]
 Distribution of the internal magnetic field depends considerably on the degree of fluid anisotropy and the sign of the parameter
 $\alpha$. Along the radius of the configuration, the magnitudes of the components of the magnetic field strength
$B_{\hat{r}}$ and $B_{\hat{\Theta}}$ may be either greater or less than those of the isotropic case.
In particular, in the case of large negative $\alpha$ they can differ in order of magnitude (see Fig.~\ref{MF_NS_fig}).

\item[(iii)] Distribution of the current is also changed considerably depending on the value of $\alpha$.
In particular, this is evident from the fact that for some values of $\alpha$ the current may have variable sign
 (see Fig.~\ref{MF_NS_fig}), which is impossible in the isotropic case.
\end{enumerate}

Therefore, we see that the structure of the internal magnetic field may depend substantially not only on the
 physical properties of matter (as demonstrated, for example, in Refs.~\cite{Bocquet:1995je,Kiuchi:2007pa} for neutron matter modeled by various EOS)
but also on the degree of fluid anisotropy.

It is clear that the obtained results are essentially model dependent and are determined by a specific manner of modeling  the anisotropy in the system.
Unfortunately,
at present there is no fully reliable way to determine the true nature of the anisotropy
and how large it may be in realistic superdense configurations.
However, if high-density matter of compact objects may possess an anisotropic pressure,
one might expect that regardless of the way in which the anisotropy is modeled, its presence will result in changes of magnetic fields,
which can be evaluated by using the approach employed here.

\section*{Acknowledgments}

We gratefully acknowledge support provided by the Volkswagen Foundation
and by a grant in fundamental research in natural sciences
of the Ministry of Education and Science of Kazakhstan.

\end{document}